\documentclass[aip,apl,reprint]{revtex4-1}

\usepackage{amsmath}
\usepackage{amssymb}
\usepackage{bm}
\usepackage{color}
\usepackage[pdftex]{graphicx}
\usepackage[latin1]{inputenc}
\usepackage[english]{babel}

\begin{document}

\preprint{AIP/123-QED}

\title{Ultrathin ferrimagnetic GdFeCo films with very low damping}

\author{Lakhan Bainsla*}
\email{lakhan.bainsla@physics.gu.se}
\affiliation{Physics Department, University of Gothenburg, 412 96 Gothenburg, Sweden.}
\author{Akash Kumar}
\affiliation{Physics Department, University of Gothenburg, 412 96 Gothenburg, Sweden.}
\author{Ahmad A.~Awad}
\affiliation{Physics Department, University of Gothenburg, 412 96 Gothenburg, Sweden.}
\author{Chunlei Wang}
\affiliation{Department of Applied Physics, KTH Royal Institute of Technology, 106 91 Stockholm, Sweden}
\author{Mohammad Zahedinejad}
\affiliation{Physics Department, University of Gothenburg, 412 96 Gothenburg, Sweden.}
\author{Nilamani Behera}
\affiliation{Physics Department, University of Gothenburg, 412 96 Gothenburg, Sweden.}
\author{Himanshu Fulara}
\affiliation{Physics Department, University of Gothenburg, 412 96 Gothenburg, Sweden.}
\author{Roman Khymyn}
\affiliation{Physics Department, University of Gothenburg, 412 96 Gothenburg, Sweden.}
\author{Afshin Houshang}
\affiliation{Physics Department, University of Gothenburg, 412 96 Gothenburg, Sweden.}
\author{Jonas Weissenrieder}
\affiliation{Department of Applied Physics, KTH Royal Institute of Technology, 106 91 Stockholm, Sweden}
\author{J. \AA kerman}
\email{johan.akerman@physics.gu.se}
\affiliation{Physics Department, University of Gothenburg, 412 96 Gothenburg, Sweden.}

\begin{abstract}

Ferromagnetic materials dominate as the magnetically active element in spintronic devices, but come with drawbacks such as large stray fields, and low operational frequencies. 
Compensated ferrimagnets provide an alternative as they combine the ultrafast magnetization dynamics of antiferromagnets with a ferromagnet-like spin-orbit-torque (SOT) behavior. However to use ferrimagnets in spintronic devices their advantageous properties must be retained also in 
ultrathin films ($t$ \textless{ 10} nm). In this study, ferrimagnetic Gd$_x$(Fe$_{87.5}$Co$_{12.5}$)$_{1-x}$ thin films in the thickness range $t$ = 2--20 nm 
were grown on high resistance Si(100) substrates and studied using broadband ferromagnetic resonance measurements at room temperature. By tuning their stoichiometry, 
a nearly compensated behavior is observed in 2 nm Gd$_x$(Fe$_{87.5}$Co$_{12.5}$)$_{1-x}$ ultrathin films for the first time, with an effective magnetization of $\mathit{M_{eff}}$ = 0.02 T and a low effective Gilbert damping constant of $\alpha$ = 
0.0078, 
comparable to the lowest values reported so far in 30 nm films. These results show great promise for the 
development of ultrafast and energy efficient ferrimagnetic spintronic devices. 

\end{abstract}

\maketitle

\section{Introduction}
Spintronic devices utilize the spin degree of freedom for data storage, information processing, and sensing~\cite{wolf2001spintronics,aakerman2005toward} 
with commercial applications such as hard drives, magnetic random access memories, and sensors. 
Besides conventional memory applications based on quasi-static operation of magnetic tunnel junctions, high frequency spintronic oscillators~\cite{demidov2014nanoconstriction, chen2016ieeeproc} have recently been demonstrated for analog computing applications such as bio-inspired neuromorphic computing~\cite{romera2018vowel, zahedinejad2020two}, logic operations, energy harvesting and Ising Machines.~\cite{houshang2020spin} For the first time, such oscillators are now used in commercial magnetic hard drives to facilitate writing to the disc.\cite{TOSHIBA.2021} The key challenges in developing such devices is to find material combinations which allow for fast operation, low-power consumption, non-volatility, and high endurance. Due to their natural spin polarization and easy manipulation, ferromagnetic materials (FM) dominate as active elements in these devices.\cite{chen2016ieeeproc} However, FMs come with drawbacks such as: (i) large magnetic stray fields affecting the operation of neighbouring devices; (ii) limited scalability of magnetic bits in memory devices; (iii) the operating frequency of spin-based oscillators limited by ferromagnetic resonance frequency, and (iv) slow synchronization of such oscillators. These shortcomings drive researchers to find more suitable materials for future spintronic devices.


Very recently, the interest in antiferromagnetic (AFM) spintronics~\cite{jungwirth2016antiferromagnetic, jungwirth2018multiple, baltz2018antiferromagnetic} increased rapidly, as AFM materials have no stray fields and can offer ultrafast spin dynamics, including AFM resonance frequencies in the THz region. It was theoretically
shown that such high-frequency excitations are possible to achieve without any applied magnetic field by injecting spin currents into AFM materials.\cite{gomonay2010spin, khymyn2017antiferromagnetic,sulymenko2017terahertz,cheng2016terahertz} Experiments have since demonstrated possible THz writing/reading capabilities.\cite{olejnik2018terahertz} 
However, the absence of a net magnetic moment in AFMs leads to difficulties in the read-out of the spin dynamics, including any microwave output signal from the AFM oscillators.\cite{khymyn2017antiferromagnetic,sulymenko2017terahertz,cheng2016terahertz} 

A possible solution is presented by ferrimagnets (FiMs), which combine the properties of FMs and AFMs. FiMs posses magnetic sub-lattices in the same way as AFMs do, but their sub-lattices are inequivalent. The magnetic sub-lattices in FiMs often consist of different magnetic ions, such as rare earth (e.g. Gd) and transition metal (e.g. Fe, Co) alloys (RE-TM) such as CoGd, and as a result, a large residual magnetization remains despite the two opposing sub-magnetizations. The temperature dependence of RE and TM sub-magnetizations in FiM can be quite different which result in magnetizations that can increase, and even change sign, with temperature \cite{stanciu2006ultrafast, ostler2011crystallographically}, in stark contrast to the non-monotonic decreasing temperature dependence for FMs and AFMs. Similar effects could also be seen by varying the composition of ferrimagnetic alloys instead of changing the temperature.\cite{kato2008compositional} In addition, the different properties of the two magnetic sub-lattices also results in two compensation points, namely the magnetization compensation point $T_m$ and the angular compensation point $T_a$. At $T_m$, the two magnetic sub-lattices cancel each other, which results in a zero net magnetic moment, while at T$_a$, their net angular momentum vanishes, as in AFMs. Therefore, at $T_a$, FiMs can have a near-THz resonance as in AFMs, while still having a net magnetic moment which can lead to strong read-out signals, including efficient microwave signal output from FiM-based oscillators \cite{lisenkov2019subterahertz}, as well as efficient control and excitation. FiMs also show high spin polarization which also make them suitable candidate for efficient magnetic tunnel junctions.\cite{JMDCoey2014MnRuGa}

\begin{figure*}
  \includegraphics[width=17cm]{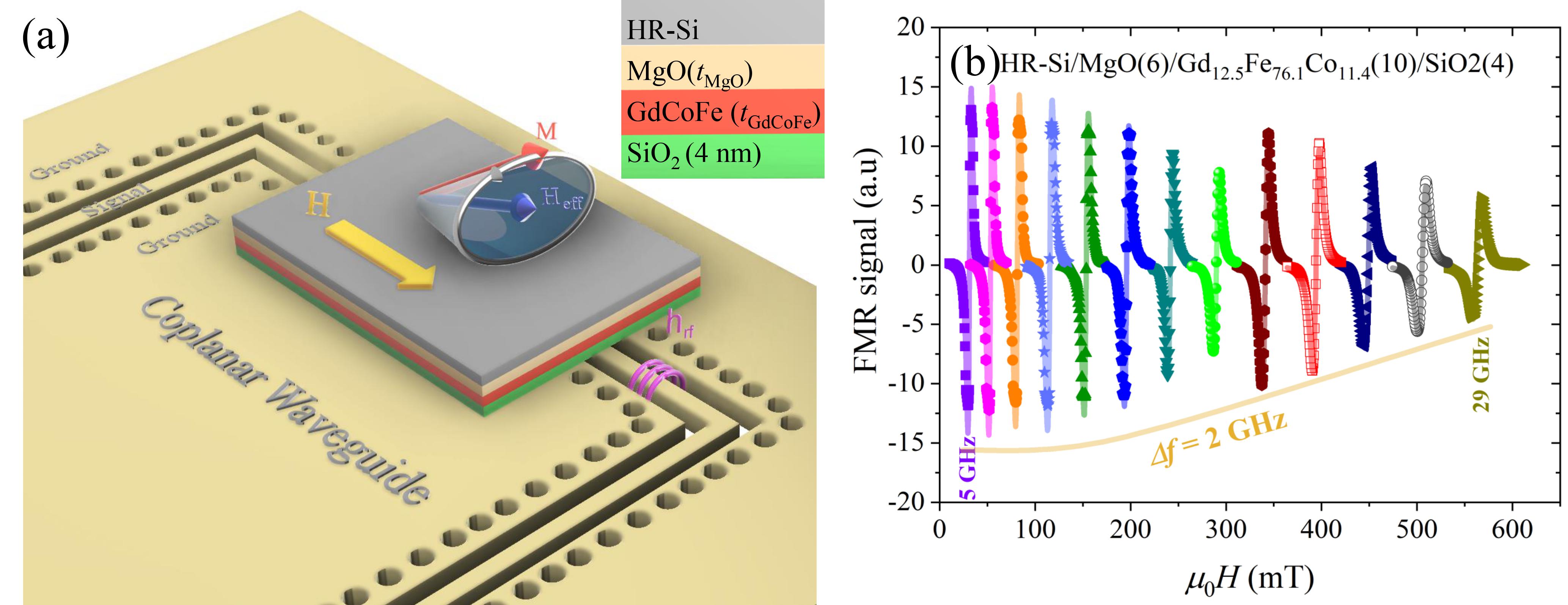}
  \centering
  \caption{(a) Schematic illustration of the coplanar waveguide (CPW), the thin film sample and its orientation, the directions of the applied magnetic field $H$, the microwave field $\mathit{h_{rf}}$, and the effective magnetic field $\mathit{H_{eff}}$ during FMR measurements. 
  Inset 
  shows the film stack. 
  (b) FMR response (derivative of the FMR absorption) for a 10 nm Gd$_{12.5}$Fe$_{76.1}$Co$_{11.4}$ film (S2) 
  recorded at different frequencies and fitted (solid lines) to Eq.~\ref{equ: FMR}. While FMR curves were recorded at 1 GHz frequency intervals throughout this study, 
  figure (b) only shows curves with $\Delta f$ = 2 GHz for clarity.}
  \label{fig1}
\end{figure*}

Due to these unique properties, research in FiMs for spintronic applications is intensifying 
\cite{finley2020spintronics}, focusing mainly on RE-TM based systems such as CoTb~\cite{finley2016spin}, CoGd~\cite{mishra2017anomalous}, and GdFeCo~\cite{roschewsky2017spin} and Mn$_{3-x}$Pt$_{x}$Ga~\cite{sahoo2016compensated,finley2019spin} based Heusler alloy. Among these, GdFeCo has been studied the most with demonstrations of fast domain wall motion~\cite{kim2017fast} and ultrafast spin dynamics \cite{stanciu2006ultrafast} near $T_a$, large spin-orbit torques and their sign reversal,~\cite{roschewsky2017spin, cespedes2021current} low magnetic damping in thick 30 nm films,~\cite{kim2019low} and sub-picosecond magnetization reversal,\cite{stanciu2007subpicosecond} to name a few. What is missing, however, is a demonstration that these unique material properties persist down to much thinner films, which will ultimately be needed if FiMs are to be used in 
spin-Hall nano oscillators (SHNOs).\cite{chen2016ieeeproc} 

In the present study, we systematically study the growth and functional properties of ultrathin ferrimagnetic Gd$_x$(Fe$_{87.5}$Co$_{12.5}$)$_{1-x}$ thin films [referred to as Gd$_x$(FeCo)$_{1-x}$ hereafter]. GdFeCo thin films in the thickness range of 2--20 nm were grown on high resistance silicon (HR-Si) substrate. The atomic composition of Gd$_x$(FeCo)$_{1-x}$ was controlled using co-sputtering and determined using inductively coupled plasma optical emission spectroscopy (ICP-OES). The magnetic properties and Gilbert damping were studied using broadband ferromagnetic resonance (FMR) measurements. 
We also demonstrate ultra low Gilbert damping for 2 nm GdFeCo, near the compensation point of Gd$_x$(FeCo)$_{1-x}$. These results paves the way for integration of FiMs into various spintronic devices and applications.

\section{Results and discussion}

The growth conditions for GdFeCo were first optimized by growing four 
10 nm thick Gd$_{12.5}$Fe$_{76.1}$Co$_{11.4}$ films on HR-Si (100) substrates using different MgO seed layer thicknesses: 0 nm (S1), 6 nm (S2), 10 nm (S3 \& S4); in S4, the seed was annealed at 600C for 1 hour prior to GdFeCo deposition to check the effect of MgO crystallinity. MgO was chosen as seed since it is insulating and therefore will not contribute any spin sinking to the magnetic damping.~\cite{spinpumpingPRL2002} 

\subsection{Seed layer dependence on 10nm thick Gd$_{12.5}$Fe$_{76.1}$Co$_{11.4}$ films}
\begin{figure*}
  \includegraphics[width=17cm]{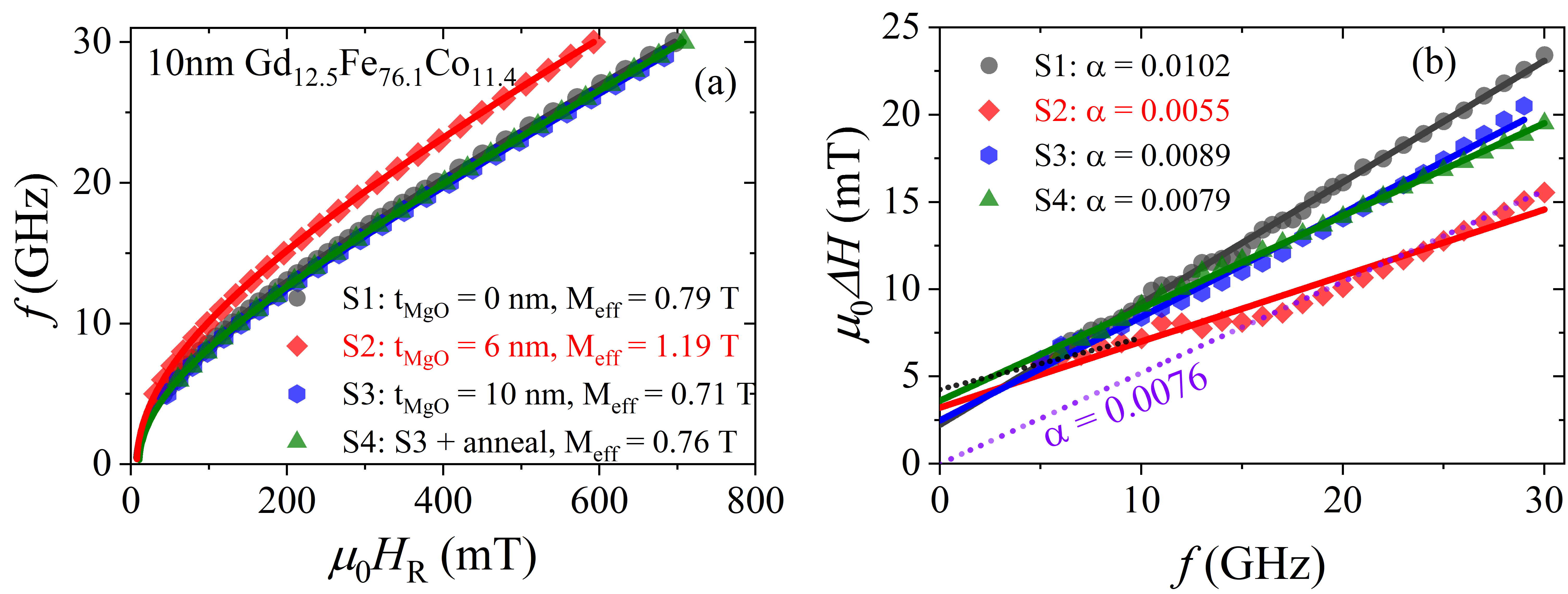}
  \caption{(a) Seed layer dependence of frequency vs resonance field of the 10 nm thick Gd$_{12.5}$Fe$_{76.1}$Co$_{11.4}$ films, here solid symbols and solid lines are the experimental data points and fitting with equation (\ref{equ: Kittel}), respectively. (b) Resonance linewidth ($\Delta{H}$) \emph{vs.}~frequency of the 10 nm thick Gd$_{12.5}$Fe$_{76.1}$Co$_{11.4}$ films, here solid symbols and solid lines are the experimental data points and fitting with equation (\ref{equ: FMRlinewidth}), respectively. The effective Gilbert damping constant values of all the samples are given in figure \ref{fig2} (b). The black and violet dotted lines in figure \ref{fig2}(b) shows the fitting of equation (\ref{equ: FMRlinewidth}) in low and high frequency regions, respectively.}
  \label{fig2}
\end{figure*}

Further details of the growth conditions are given in the experimental section. FMR measurements, on $6\times3$ mm$^2$ rectangular pieces cut from these films, were then performed using a NanOsc PhaseFMR-40 FMR Spectrometer. The sample orientation on the coplanar waveguide (CPW), together with 
the directions of the applied field, the microwave excitation field $\mathit{h_{rf}}$, and the effective magnetic field $\mathit{H_{eff}}$, are shown in Fig.~\ref{fig1}(a). 
Typical (derivative) FMR absorption spectra obtained for S2 are shown in figure \ref{fig1}(b) together with fits to a 
sum of symmetric and anti-symmetric Lorentzian derivatives:~\cite{woltersdorf2004spin} 
\begin{equation}\label{equ: FMR}
   \frac{dP}{dH}(H)=\frac{-8C_1\Delta{H}(H-H_R)}{[\Delta{H}^2+4(H-H_R)^2]^2}+\frac{2C_2(\Delta{H}^2-4(H-H_R)^2)}{[\Delta{H}^2+4(H-H_R)^2]^2}
\end{equation}
where $H_R$, $\Delta{H}$, $C_1$, and $C_2$ represent the resonance field, the full width at half maximum (FWHM) of the FMR absorption, and the symmetric and anti-symmetric fitting parameters of the Lorentzian derivatives, respectively. 
The extracted values of $H_R$ \emph{vs.}~$f$ 
are shown in figure \ref{fig2} (b) together with fits to Kittel's equation:\cite{Kittelfmr}
\begin{equation}\label{equ: Kittel}
   f=\frac{\gamma\mu_0}{2\pi}\sqrt{(H_{R}-H_{k})(H_{R}-H_{k}+M_{eff})}
\end{equation}
where, $\gamma$, $H_k$ and $\mathit{M_{eff}}$ are the gyromagnetic ratio, the in-plane magnetic anisotropy field, and the effective magnetization of the sample, respectively, all allowed to be free fitting parameters. Values for $\gamma$ and $H_k$ only showed minor variation between the four samples, with $\gamma/2\pi$ = 29.4-30.0 GHz/T and $H_k$ = 66-104 Oe. 
$\mathit{M_{eff}}$ varied more strongly, with values of 0.79, 1.19, 0.71 and 0.76 T obtained for S1, S2, S3 and S4, respectively. 

\begin{figure*}
  \includegraphics[width=17cm]{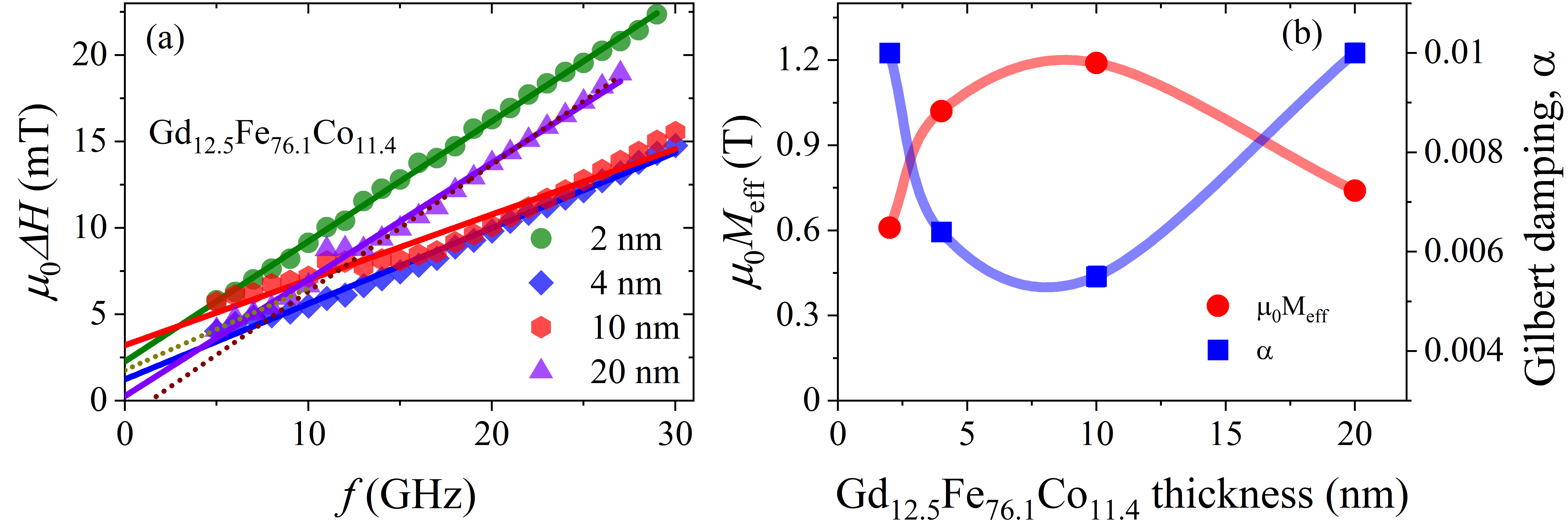}
  \caption{(a) FMR linewidth $\Delta{H}$ \emph{vs.}~$f$ for four Gd$_{12.5}$Fe$_{76.1}$Co$_{11.4}$ films with different thicknesses, together with linear fits to 
  equation (\ref{equ: FMRlinewidth}). 
  The dotted lines 
  show fits 
  for the 20 nm film in its low and high frequency regions, respectively. (b) Effective magnetization and effective Gilbert damping constant \emph{vs.}~
  thickness; lines are guides to the eye.} 
  \label{fig3}
\end{figure*}

The effective Gilbert damping constant $\alpha$ can then be obtained from fits of $\Delta H$ \emph{vs.}~$f$ to:~\cite{bainsla2018low} 
\begin{equation}\label{equ: FMRlinewidth}
  \Delta H=\Delta H_{0}+\frac{4\pi\alpha f}{\gamma\mu_0}
\end{equation}
where the offset $\Delta{H}_0$ represents the inhomogeneous broadening. Equation (\ref{equ: FMRlinewidth}) is well fitted to the experimental values, using $\Delta{H}_0$ and $\alpha$ as adjustable fitting parameters for all the four samples, as shown in the figure \ref{fig2}(b). $\Delta{H}_0$ = 2--4 mT is essentially sample independent within the measurement accuracy. In contrast, the obtained values of $\alpha$ vary quite strongly and are given inside figure \ref{fig2}(b). The GdFeCo grown with 6 nm MgO seed layer (S2) clearly shows the lowest value of ${\alpha}=0.0055$, although this might be affected by the slight non-linear behavior around 10 to 15 GHz. However, when only the high-field data is fitted, the extracted damping of ${\alpha}$ = 0.0076 is still the lowest and at all frequencies the linewidth of S2 lies well below all the other samples. As damping is one of the most important parameters for spintronic devices, we hence chose the growth conditions of S2 for all subsequent films in this study.

\subsection{Thickness dependence on Gd$_{12.5}$Fe$_{76.1}$Co$_{11.4}$ films}
After optimizing the growth conditions for Gd$_{12.5}$Fe$_{76.1}$Co$_{11.4}$, the thickness dependence of the films was studied with the same composition using the growth conditions of sample S2. The 
FMR linewidth $\Delta{H}$ \textit{vs.~f} is shown in figure \ref{fig3}(a) and exhibits a relatively strong dependence on thickness. 
It is noteworthy that the 4 nm film shows the narrowest linewidth at all frequencies, clearly demonstrating that very low damping can be achieved also in ultra-thin GdFeCo. The extracted $\mathit{M_{eff}}$ and $\alpha$ are shown \emph{vs.}~thickness in figure \ref{fig3}(b), both showing a strong thickness dependence. 
Damping as low as ${\alpha}=0.0055$ is obtained for the 10 nm thick films. If only the high-field portion of the data is fitted, the extracted damping increases to 
0.0076, which is still about an order of magnitude lower than any literature value on 10 or 30 nm films.\cite{okuno2019temperature,kato2008compositional} Both the 10 and 20 nm films 
showed a minor nonlinearity in $\Delta{H}$ \emph{vs.}~$f$ data and were therefore analysed by fitting the data in both the low and the high field regions separately, as shown by the dotted lines in figure \ref{fig3}(a). The $\alpha$ value for the 20 nm film increased slightly from 0.0098 to 0.0109 if only high field data is used for analysis. The relatively higher damping for the 20 nm film might be due to the radiative damping mechanism which increases proportionally with magnetic layer thickness.~\cite{schoen2016ultra} We conclude that 2 nm ultrathin films can indeed be grown with reasonably low damping. Since the damping is strongly thickness dependent in this regime, the optimum thickness for devices may likely be found in the 2--4 nm range.


\begin{figure*}
  \includegraphics[width=14cm]{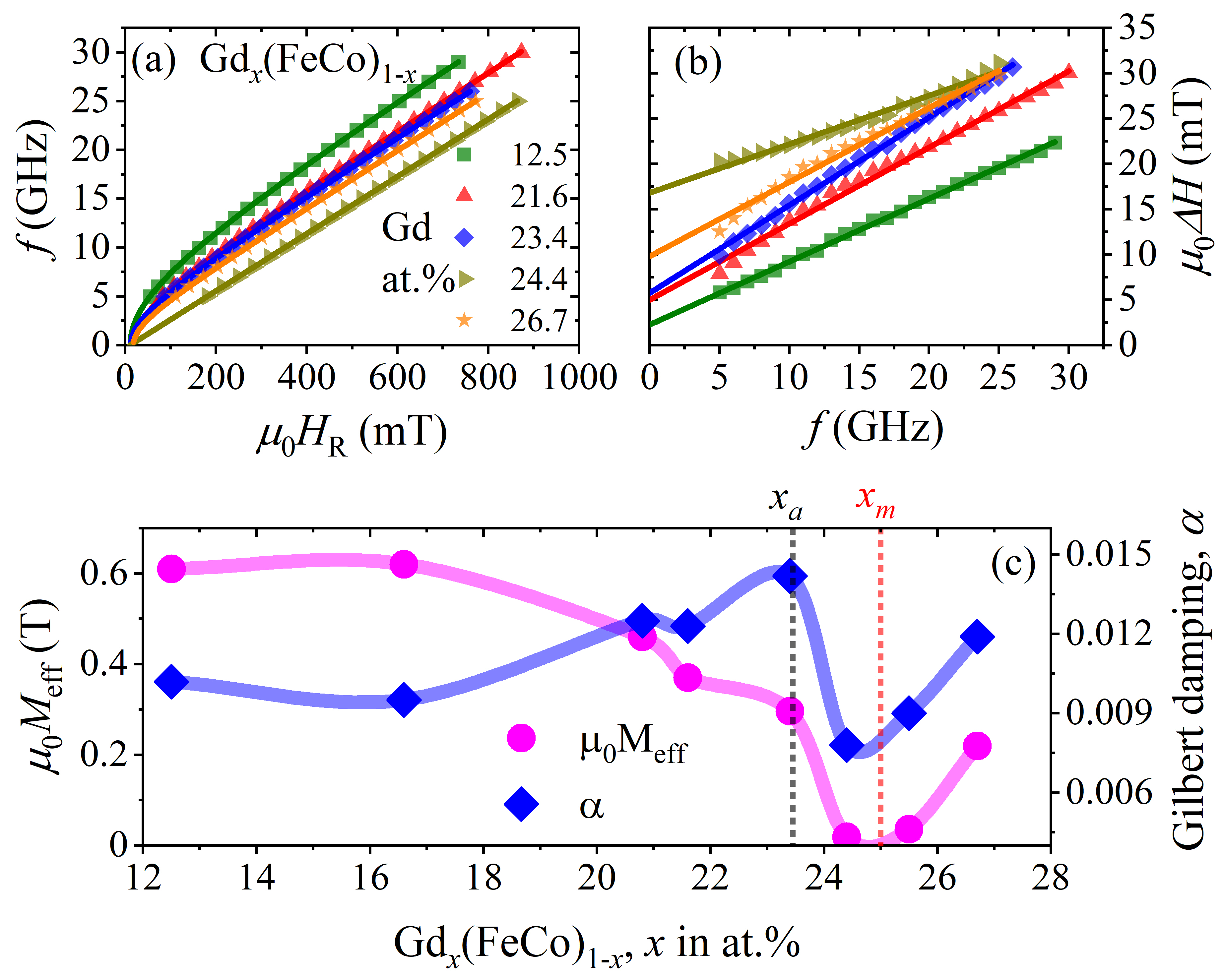}
  \caption{(a) Frequency \emph{vs.}~resonance field and (b) resonance linewidth \emph{vs.}~frequency, of 2 nm thick Gd$_x$(FeCo)$_{1-x}$ films as a function of 
  Gd content in atomic \%. 
  (c) Effective magnetization and effective Gilbert damping constant \emph{vs.}~Gd content. 
  Solid symbols represent the values obtained by fitting the experimental FMR data in (a) and (b) using the equation (\ref{equ: Kittel}) and (\ref{equ: FMRlinewidth}), respectively; solid lines in (c) are guides to the eye. 
  $\textit{x}_a$ and $\textit{x}_m$ show the angular and magnetic compensation points, respectively, obtained from the literature~\cite{stanciu2006ultrafast, kato2008compositional}.} 
  \label{fig4}
\end{figure*}

\subsection{Composition dependence on 2nm thick films}
To finally investigate whether we can achieve a 
compensated ferrimagnetic behavior also in ultra-thin films, we grew 
2 nm Gd$_x$(FeCo)$_{1-x}$ films in the composition range 12--27 at.\% Gd. 
The films were characterized using FMR spectrometry as described above and the extracted results are shown in figure \ref{fig4}. 

\begin{table*}
 \caption{The obtained values of effective Gilbert damping constant $\alpha$ at room temperature (RT) in this work and comparison with the lowest values reported so far in the literature at RT and also at their respective angular momentum compensation (T$_a$) and magnetic compensation (T$_m$) points.}
  \begin{center}
  \begin{tabular}{@{}cccccc@{}}
    \hline
    Film composition & Film thickness & $\alpha$ & Measurement technique & Analysis method & Reference \\
    \hline
    Gd$_{23.5}$Fe$_{68.9}$Co$_{7.6}$ & 30  & $\sim$ 0.45 (at RT) & FMR  & Kittel's FMR  & \cite{kato2008compositional}  \\
      & & $\sim$ 0.35 (at RT) & Pump-probe & &  \\
      &  &  &  &  & \\
    Gd$_{22}$Fe$_{74.6}$Co$_{3.4}$ & 20  & $\sim$ 0.21 (at T$_a$) & Pump-probe  & -do-  & \cite{stanciu2006ultrafast}  \\
      & & $\sim$ 0.13 (at T$_m$) & & &  \\
      &  &  &  &  & \\
    Gd$_{25}$Fe$_{65.6}$Co$_{9.4}$ & 10  & $\sim$ 0.07 (at RT) & Spin torque FMR  & -do-  & \cite{okuno2019temperature}  \\
      & & $\approx$ 0.01 (at RT) & Spin torque FMR & Ferrimagnetc resonance &  \\
      &  &  &  &  & \\
    Gd$_{23.5}$Fe$_{66.9}$Co$_{9.6}$ & 30  & 0.0072 (at RT) & Domain wall (DW)  & Field driven DW & \cite{kim2019low}\\
     &  & & motion  & mobility & \\
    &  &  &  &  & \\
    Gd$_{12.5}$Fe$_{76.1}$Co$_{11.4}$ & 10  & 0.0055 & Broadband FMR  &  Kittel's FMR & This work \\
    &  & 0.0076 (HF data) & -do- & -do-  & This work \\
    &  &  &  &  & \\
    Gd$_{12.5}$Fe$_{76.1}$Co$_{11.4}$ & 4  & 0.0064 & -do-  &  -do- & This work \\
    &  &  &  &  & \\
    Gd$_{12.5}$Fe$_{76.1}$Co$_{11.4}$ & 2  & 0.0101 & -do-  &  -do- & This work \\
    &  &  &  &  & \\
    Gd$_{23.4}$Fe$_{67.0}$Co$_{9.6}$ & 2  & 0.0141 & -do-  &  -do- & This work \\
    &  &  &  &  & \\
    Gd$_{24.4}$Fe$_{66.1}$Co$_{9.5}$ & 2  & 0.0078 & -do-  &  -do- & This work \\
    \hline
  \end{tabular}
  \end{center}
\end{table*} 

The extracted $\mathit{M_{eff}}$ and $\alpha$ 
follow a similar trend as reported earlier for one order of magnitude thicker GdFeCo films characterized using an all-optical pump-probe technique.~\cite{stanciu2006ultrafast} We first note that we can indeed reach an essentially fully compensated antiferromagnetic behavior in two films around a composition of 25 at.$\%$ Gd. We have marked this compensation point with $x_m$ and a dashed line in figure \ref{fig4} (c). Both films show very low damping of 0.0078 and 0.009 respectively. However, just below this composition, the damping shows a peak, which is consistent with an \emph{angular} compensation point, which we denote by $x_a$. It is noteworthy that the extracted damping value of $\alpha$ = 0.0142 is still more than an order of magnitude lower than $\alpha$ = 0.45 of 30 nm films measured using FMR spectrometry~\cite{kato2008compositional} and $\alpha$ = 0.20 of 20 nm films measured using an optical pump-probe technique.~\cite{stanciu2006ultrafast}  

\section{Conclusion}
In view of the potential application of compensated ferrimagnets to spintronic devices, we prepared ferrimagnetic thin films of Gd$_x$(FeCo)$_{1-x}$ on high resistance Si(100) substrates and studied them using the FMR measurements. Their growth conditions were optimized using 10 nm thick Gd$_{12.5}$Fe$_{76.1}$Co$_{11.4}$ films, after which thickness dependent studies were done on the same composition in the thickness range of 2--20 nm. Composition dependence studies were finally done on 2 nm thick Gd$_x$(FeCo)$_{1-x}$ films and an essentially compensated ferrimagnetic behavior was observed for the first time in ultrathin 2 nm films. The angular momentum compensation and magnetic compensation points observed in this work are very close to those reported earlier on much thicker films in the literature. A record low  $\alpha$ value of about 0.0078 is obtained near the magnetic compensation point, which is an order of magnitude lower than the values reported in the literature using similar analysis methods. The observation of compensated ferrimagnetic behavior in ultrathin films together with very low value of $\alpha$ are promising  results for the future development of ultrafast and energy efficient ferrimagnetic spintronic devices.


\section*{Experimental Section}
\subsection{Thin films growth and composition analysis}All the samples were prepared on high resistivity Si(100) substrates using a magnetron sputtering system with a base pressure of less than $2\times10^{-8}$ torr. Thin films of Gd$_x$(FeCo)$_{1-x}$ were deposited using the co-sputtering of high purity (more than 99.95\%) Gd and Fe$_{87.5}$Co$_{12.5}$ targets, and composition analysis was done using the inductively coupled plasma mass spectroscopy (ICP-MS). Thin films stacking structure of Si(100)/MgO(t)/Gd$_{12.5}$Fe$_{76.1}$Co$_{11.4}$(10)/SiO$_2$(4) were used for seed layer dependence studies, here, the number in the bracket is the thickness of the layer in nm, where \textit{t} =0, 6 and 10 nm. Four samples, namely S1 to S4 were prepared to obtain the best conditions to grow Gd$_{12.5}$Fe$_{76.1}$Co$_{11.4}$(10) films. For S1, Gd$_{12.5}$Fe$_{76.1}$Co$_{11.4}$(10) was grown directly over HR-Si (100) substrates, while in both S2 and S3 Gd$_{12.5}$Fe$_{76.1}$Co$_{11.4}$ were grown with MgO seed layer of 6 and 10 nm, respectively. All the layers in S1-S3 were grown at room temperature and no further heat treatment was given to them. In S4, 10 nm MgO seed layer were grown over HR Si(100) substrates at RT and followed by a in-situ post-annealing at 600C for 1 hour, and after that Gd$_{12.5}$Fe$_{76.1}$Co$_{11.4}$ were deposited.
The stacking structure of Si(100)/MgO(6)/Gd$_{12.5}$Fe$_{76.1}$Co$_{11.4}$(m)/SiO$_2$(4) were used for thickness dependence studies, where m is the thickness of Gd$_{12.5}$Fe$_{76.1}$Co$_{11.4}$ layer, and varied from 2 to 20 nm. For composition dependence studies, stacking structure of Si(100)/MgO(6)/Gd$_{x}$(FeCo)$_{1-x}$(2)/SiO$_{2}$(4) were used, where \textit{x} varied from 12.5 to 26.7. The composition of Gd$_{x}$(FeCo)$_{1-x}$ films was varied by changing the sputtering rate of Fe$_{87.5}$Co$_{12.5}$ target, while keeping the Gd sputtering rate fixed for most films. All the samples for thickness dependence and composition dependence were grown at room temperature and no post-annealing was used. Layer thicknesses were determined by estimating the growth rate using the Dektak profiler on more than 100 nm thick films.\par
\subsection{Inductively coupled plasma mass spectroscopy (ICP-MS) measurements}
The elemental composition (Co, Fe, and Gd) of the thin film samples was determined by inductively coupled plasma optical emission spectroscopy (ICP-OES) using a Thermo Fisher Scientific iCAP 6000 Series spectrometer. Each thin film sample was exhaustively extracted in 5 mL HNO3 (65$\%$, Supelco, Merck KgaA, Sigma-Aldrich) for a duration of 30 min. 5 mL ultrapure MilliQ-water (18 M$\Omega$cm) was added to the solution and the extract was allowed to rest for 30 minutes.  The extract was transferred to a 100 mL volumetric flask. The extracted sample was then rinsed for several cycles in ultrapure water. The water used for rinsing was transferred to the same volumetric flask. The extract was diluted to 100 mL for ICP analysis. ICP check standards were prepared from standard solutions (Co and Fe: Merck, Germany; Ga: Accustandard, USA). The relative standard deviation (from three individual injections) were within 1$\%$.
\par

\subsection{Ferromagnetic resonance (FMR) measurements}Rectangular pieces of about 6$\times$3 mm$^{2}$ were cut from the blanket films and broadband FMR spectroscopy was performed using a NanOsc Phase FMR (40 GHz) system with a co-planar waveguide for microwave field excitation. Microwave excitation fields $\mathit{h_{rf}}$ with frequencies up to 30 GHz were applied in the film plane, and perpendicular to the applied in-plane dc magnetic field H. All the FMR measurements were performed at the room temperature. The schematic of FMR measurement setup is shown in \ref{fig1}(a), and further details about the measurements are given in Section 2 (results and discussions).\par

\section*{Supporting Information} 
Supporting Information is available from the Wiley Online Library or from the corresponding author.\\

\section*{Acknowledgements}
 Lakhan Bainsla thanks MSCA - European Commission for Marie Curie Individual Fellowship (MSCA-IF Grant No. 896307). This work was also partially supported by the Swedish Research Council (VR Grant No. 2016-05980) and the Horizon 2020 research and innovation programme (ERC Advanced Grant No.~835068 "TOPSPIN"). 

\section*{Conflict of Interest}
 The authors declare no conflict of interest. 

\section*{Author Contributions}
  L.B. and J.\AA. planned the study. L.B. grew the films, performed the FMR measurements and analysed the obtained FMR data. J.W. helped with ICP-MS measurements and analysis. L.B. wrote the original draft of the paper. J.\AA. coordinated and supervised the work. All authors contributed to the data analysis and co-wrote the manuscript.

\section*{Data Availability Statement}
  The data that support the findings of this study are available from the corresponding author on reasonable request.\\

\medskip

\bibliographystyle{apsrev4-1}
\bibliography{Main.bib}

\end{document}